\documentclass{article}

\usepackage[nonatbib,final]{neurips_2020_ml4ps}

\usepackage[utf8]{inputenc} 
\usepackage[T1]{fontenc}    
\usepackage{hyperref}       
\usepackage{url}            
\usepackage{booktabs}       
\usepackage{amsfonts}       
\usepackage{nicefrac}       
\usepackage{microtype}      
\usepackage{xcolor}
\usepackage{mwe}

\title{The Mass-ive Issue: Anomaly Detection in Jet Physics}

\author{%
    Tobias Golling, Takuya Nobe, Dimitrios Proios, John Andrew Raine\\ \textbf{Debajyoti Sengupta, Slava Voloshynovskiy}\\
    Universit\'e de Gen\`eve\\
    \texttt{tobias.golling@unige.ch, tnobe@cern.ch, dimitrios.proios@etu.unige.ch}\\ \texttt{jraine@cern.ch, desengup@cern.ch, svolos@unige.ch}
    \AND
    Jean-Francois Arguin, Julien Leissner Martin, Jacinthe Pilette\\
    Universit\'e de Montreal \\
    \texttt{arguin@lps.montreal.ca, julien.leissner-martin@umontreal.ca}\\ \texttt{jacinthe.pilette@umontreal.ca}
    \AND
    Debottam Bakshi Gupta, Amir Farbin\\
    University of Texas, Arlington\\
    \texttt{dbakshig@cern.ch, afarbin@cern.ch}
}

\begin{document}

\maketitle


\begin{abstract}
    In the hunt for new and unobserved phenomena in particle physics, attention has turned in recent years to using advanced machine learning techniques for model independent searches. In this paper we highlight the main challenge of applying anomaly detection to jet physics, where preserving an unbiased estimator of the jet mass remains a critical piece of any model independent search. Using Variational Autoencoders and multiple industry-standard anomaly detection metrics, we demonstrate the unavoidable nature of this problem.
\end{abstract}

\section{Introduction}

An area of extreme importance in particle physics is the hunt for new physics, observing phenomena outside of the predictions of the Standard Model (SM) of particle physics. The SM has been the established theory of fundamental particles and their interactions for over forty years, however, in order to completely describe nature we know that there are areas in which it is not a complete model. Attention therefore turns to the hunt for new processes that could explain these discrepancies and manifest in the existence of new fundamental particles.

The Large Hadron Collider (LHC) provides a perfect window to search for signatures left behind by these particles, that are expected to be produced in collisions. A standard signature that can be expected to be observed in the general purpose detectors at the LHC is the presence of jets, a collimated shower of energy deposited by particles in the detector, initiated by a single particle and its decay processes and subsequent shower.



The major background in model independent searches comes from jets initiated by quarks and gluons (QCD jets), fundamental particles with negligible mass. The observed mass of QCD jets is a falling spectrum across all energies, in comparison to a peak centred around the particle mass observed for jets initiated by massive particles. This background is incredibly dominant, and being able to identify a potential signal from the mass alone is almost impossible.
Therefore, we employ anomaly detection techniques to identify jets which are unlike QCD jets, enhancing the presence of these signals and making their mass peaks visible, without making strong assumptions on the underlying physics that would need to be made in a supervised setting.


\section{Inputs, Architecture and Metrics}

\begin{figure}[h]
    \centering
    \includegraphics[width=0.66\textwidth]{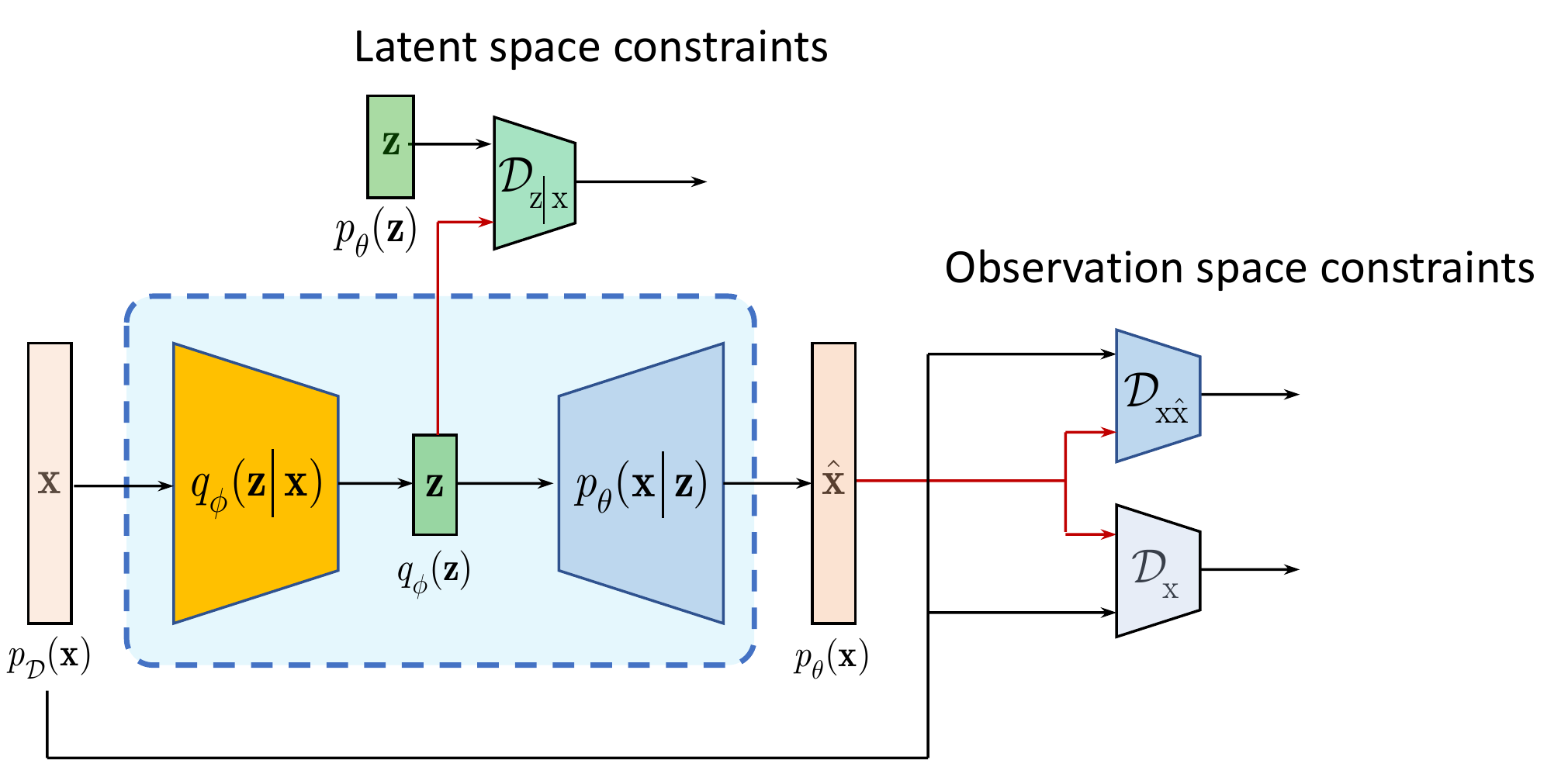}
    \caption{Model architecture used for jet reconstruction and anomaly detection.
    The autoencoder networks are highlighted in the blue dashed box, with the input and reconstructed output, $\mathrm{x}$ and $\mathrm{\hat{x}}$, shown in light red. The latent space distribution, $\mathrm{z}$, is represented by a green rectangle for both the true distribution $p_\theta (\mathrm{z})$ and the distribution sampled from the encoded mean and variance from the encoder $q_\phi (\mathrm{z})$. The connections between the input, output and latent space to the three constraints, which are used as loss terms in the training, are depicted by arrows.}
    \label{fig:VAE}
\end{figure}

In this paper we employ Variational Autoencoders (VAEs)~\cite{VAE,betaVAE} motivated by the Information Bottleneck theory~\cite{tishby2015dee} extended to unsupervised setting \cite{Voloshynovskiy2019NeurIPS} to reconstruct QCD jets. The architecture of the trained model is shown in Fig.~\ref{fig:VAE}.
We simulate jets from proton-proton collisions at a centre of mass energy $\sqrt{s}=13$~TeV using MadGraph+Pythia8~\cite{Madgraph,Pythia} interfaced with Delphes~\cite{Delphes} to simulate the interactions in the ATLAS detector~\cite{ATLAS} at the LHC. Jets are clustered using the anti-$k_{\mathrm{T}}$ algorithm~\cite{antikt} with a radius parameter $\Delta R=1.0$ using particle flow objects and without applying any trimming. Events were simulated
without the inclusion of pile-up interactions. A minimum jet $p_\mathrm{T}$ cut of 450~GeV is applied on the leading jet of a simulated event, emulating the large jet trigger threshold of the ATLAS detector.

In this study we use simulated QCD jets to train the VAE. To test its performance for anomaly detection we simulate jets initiated by top quarks decaying hadronically, and two hypothetical new physics models based on Stealth Bosons~\cite{StealthBosonJA}, one with a mass of 80 GeV (\textit{StealthB(80)}), and another with a mass of 400 GeV (\textit{StealthB(400)}). Each Stealth Boson decays in a four prong process via two pseudo-scalars of mass 30~GeV and 80~GeV respectively.

\begin{figure}[h]
    \centering
    \includegraphics[width=0.95\textwidth]{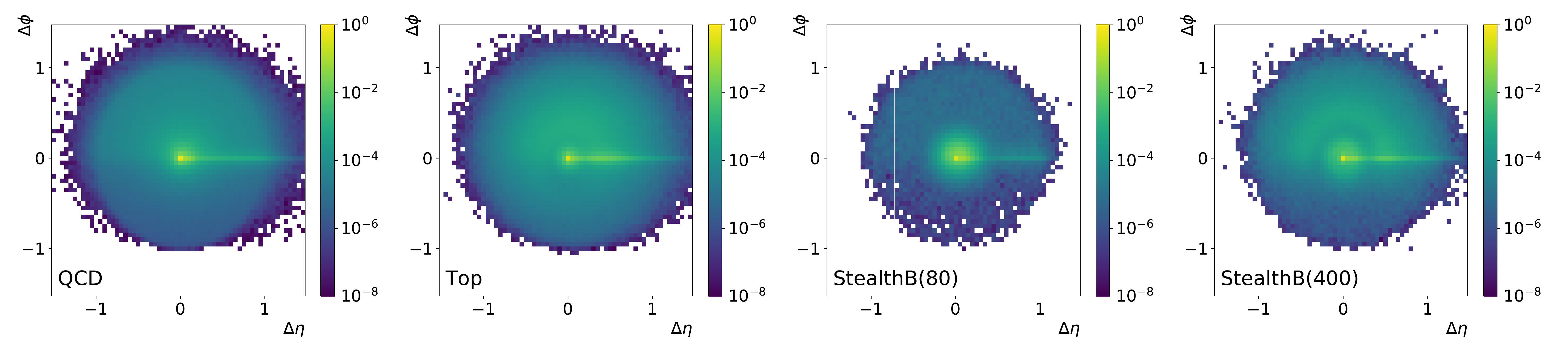}
    \caption{Average distribution of momentum within a jet in the $\eta-\phi$ plane for the four particles considered in this paper after applying preprocessing. From left to right: QCD, Top, StealthB(80) and StealthB(400). The intensity corresponds to $p_{\mathrm{T}}$ fraction of the jet.}
    \label{fig:jetimg}
\end{figure}

As an input to the VAE we use up to the 20 most energetic constituents of each jet, with zero padding in the case of fewer constituents.
It was found that for a higher number of constituents the reconstruction performance and ability to detect anomalies did not improve.
We represent these constituents as a four-momenta point cloud, with each point in the form $(p_x,p_y,p_z,E)$. As each constituent is treated as massless, the $E$ term is the quadratic sum of the three momenta. A jet is reconstructed as the linear sum of all constituents.
In order to reduce ambiguity from degrees of freedom, a preprocessing involving centering and rotating the jet constituents is applied, following the procedure proposed in Ref.~\cite{JetImages}. Furthermore, each component is normalised by the transverse momentum of the jet $(p_{\mathrm{T}})$. Because only the leading 20 constituents are considered, the sum of constituent $p_\mathrm{T}$ can be less than the jet $p_\mathrm{T}$, and the reconstructed mass can be lower than predicted.
The average signature of these jets is shown in Fig.\ref{fig:jetimg} for two known processes, and two hypothetical particles. It should be noted that in contrast to the average images, an individual jet is stochastic and sparse, which presents additional challenges.
The constituents are flattened into an 80-dimensional vector with a linear scaling applied to all input dimensions, using the median and interquartile range in each of the input dimensions separately.

The model used for the studies in this paper is shown in Fig.~\ref{fig:VAE}. 
The model is based on the variational decomposition of unsupervised information bottleneck considered in Ref.~\cite{Voloshynovskiy2019NeurIPS}. The first term of this decomposition corresponds to the mutual information minimization between the input and latent space. It is represented by an upper bound, which also used in the traditional VAE, and is denoted as $\mathcal{D}_{\mathrm{z}|\mathrm{x}}$. The second term is responsible for the maximisation of the mutual information between the latent space and the reconstruction space. It has an upper bound set by a conditional entropy term $\mathcal{D}_{x\hat{x}}$, which ensures the accurate reconstruction on a sample level, and the Kullback-Leibler divergence term $\mathcal{D}_\mathrm{x}$, which ensures a correspondence between the total distributions of reconstructed and input data.
On the data set presented in this paper, the addition of $\mathcal{D}_\mathrm{x}$ dramatically improves the reconstruction fidelity of the average jet shower, visible when looking at the representation shown in Fig.~\ref{fig:jetimg} for reconstructed jets. When considering only the $\mathcal{D}_{x\hat{x}}$ term in the training, the relationship between the average $p_\mathrm{T}$ fraction of constituents, which span several orders of magnitude, and the location of a constituent is not accurately preserved, and a smeared central region is seen in the average representation.

The encoder and decoder (mirrored) consist of fully connected layers of 256, 128 and 64 nodes. The latent space has 10 dimensions, with $\mathrm{z}$ sampled from the encoded mean and variance. This architecture is chosen to match the network presented Ref.~\cite{oevae}, which is seen to have good reconstruction performance.
A Gaussian prior constraint on  $\mathrm{z}$ is enforced by the Kullback-Leibler divergence term, denoted as $\mathcal{D}_{\mathrm{z}|\mathrm{x}}$, with a loss weight of 0.25. The two observation space constraints are the mean square error (MSE, here denoted as $\mathcal{D}_{x\hat{x}}$) between the input and output, and an adversarial discriminator ($\mathcal{D}_\mathrm{x}$) trained to identify the input from reconstructed outputs. $\mathcal{D}_\mathrm{x}$ is connected to the output of the network via a gradient reversal layer~\cite{GradReversal}. A gradient penalty of 10 with the loss term multiplied by 0.1 is used to lower the effective learning rate of the discriminator in comparison to the variational autoencoder.
The network is trained without $\mathcal{D}_\mathrm{x}$ for an initial 50 epochs, at a learning rate of $10^{-3}$. This is followed by 10 epochs with the adversarial network included in the training, with all three loss terms and a learning rate of $10^{-5}$. Finally, the network is trained for a further 90 epochs with the weights in $\mathcal{D}_\mathrm{x}$ frozen, using all three loss terms and a learning rate of $10^{-3}$.
The model is implemented and trained using TensorFlow~2.2~\cite{Tensorflow}.
The discriminator network is frozen to ensure a weak discriminator which was found to perform better for anomaly detection.

More complicated architectures were tested in these studies, for example including the use of a recursive input instead of a flat vector, and an adversarial discriminant for the latent space prior. However, all networks presented the same behaviour as the model presented in this paper when applied to anomaly detection. The final choice of the hyperparameters used in training was made based on the reconstruction of the jet mass and the ability of the $\mathcal{D}_\mathrm{x}$ network to separate QCD jets from top jets for both the input and reconstructed jets.

In this paper five metrics to detect anomalous jets are compared. Firstly, the \textit{MSE} of a jet passed through the network is considered. A second observation space metric called the \textit{Energy Movers Distance}~(EMD)~\cite{EMD}, motivated by optimal transport problems, is employed between an input jet and its reconstructed output using the momenta and energy as a fraction of jet $p_\mathrm{T}$ for the leading 20 constituents.
Two approaches are employed in the latent representation of a jet, using the encoded means. Firstly, a \textit{$k$-Nearest Neighbour} (kNN) algorithm~\cite{sklearn} is trained on the encoded latent space with the sum distance of the nearest neighbours to the encoded point of a jet used as a metric. Secondly, a \textit{One Class Support Vector Machine}~(OCSVM)~\cite{sklearn} is trained on the encoded latent representation. In both cases, only the class of jets used for training the network are used to train the latent space metrics.
Finally, the output of $\mathcal{D}_\mathrm{x}$ for the jet in the input space~(Dx(in)) is studied.

\section{Performance}

\begin{figure}[h]
    \centering
    \includegraphics[width=0.45\textwidth]{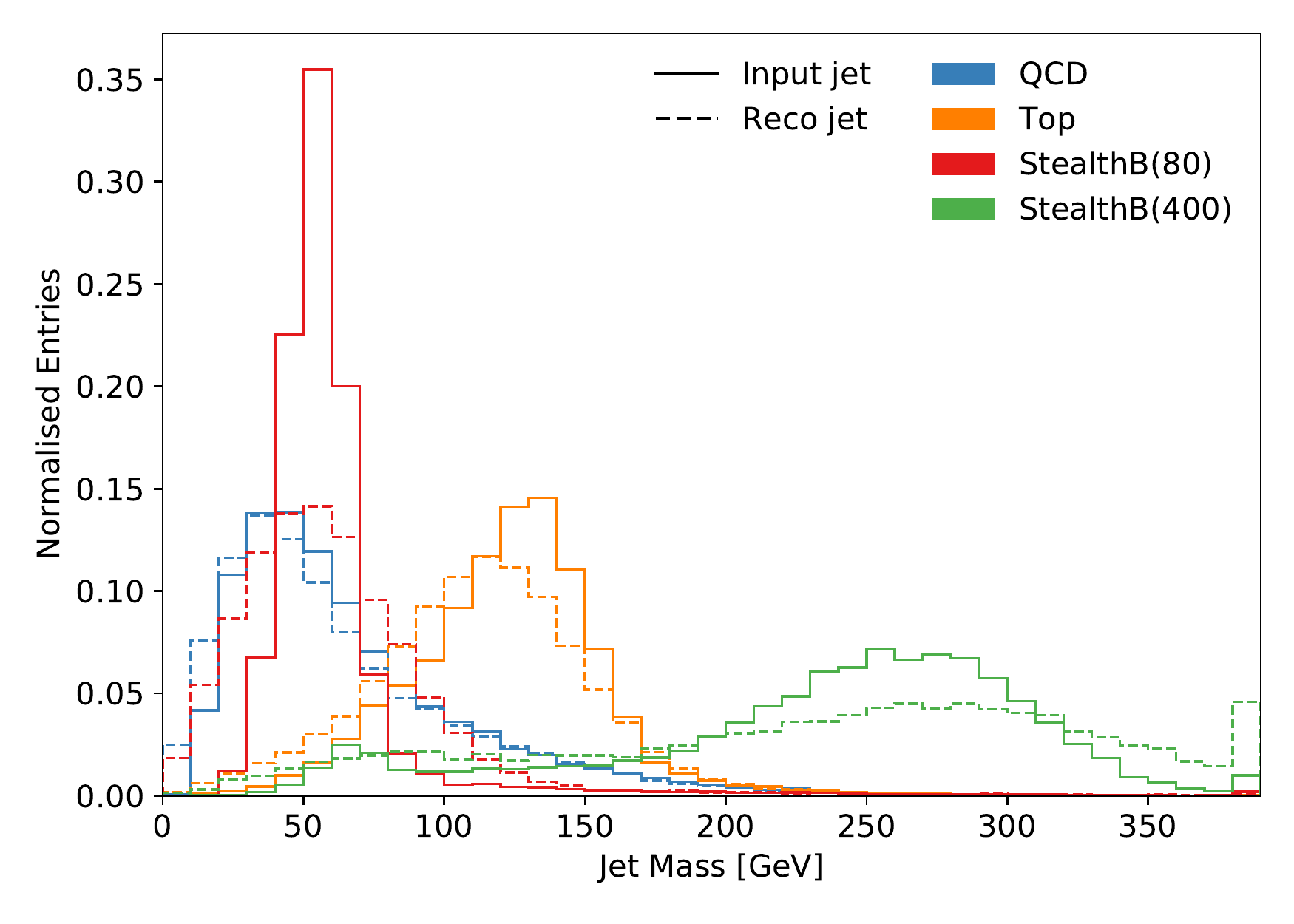} \hskip4ex
    \includegraphics[width=0.45\textwidth]{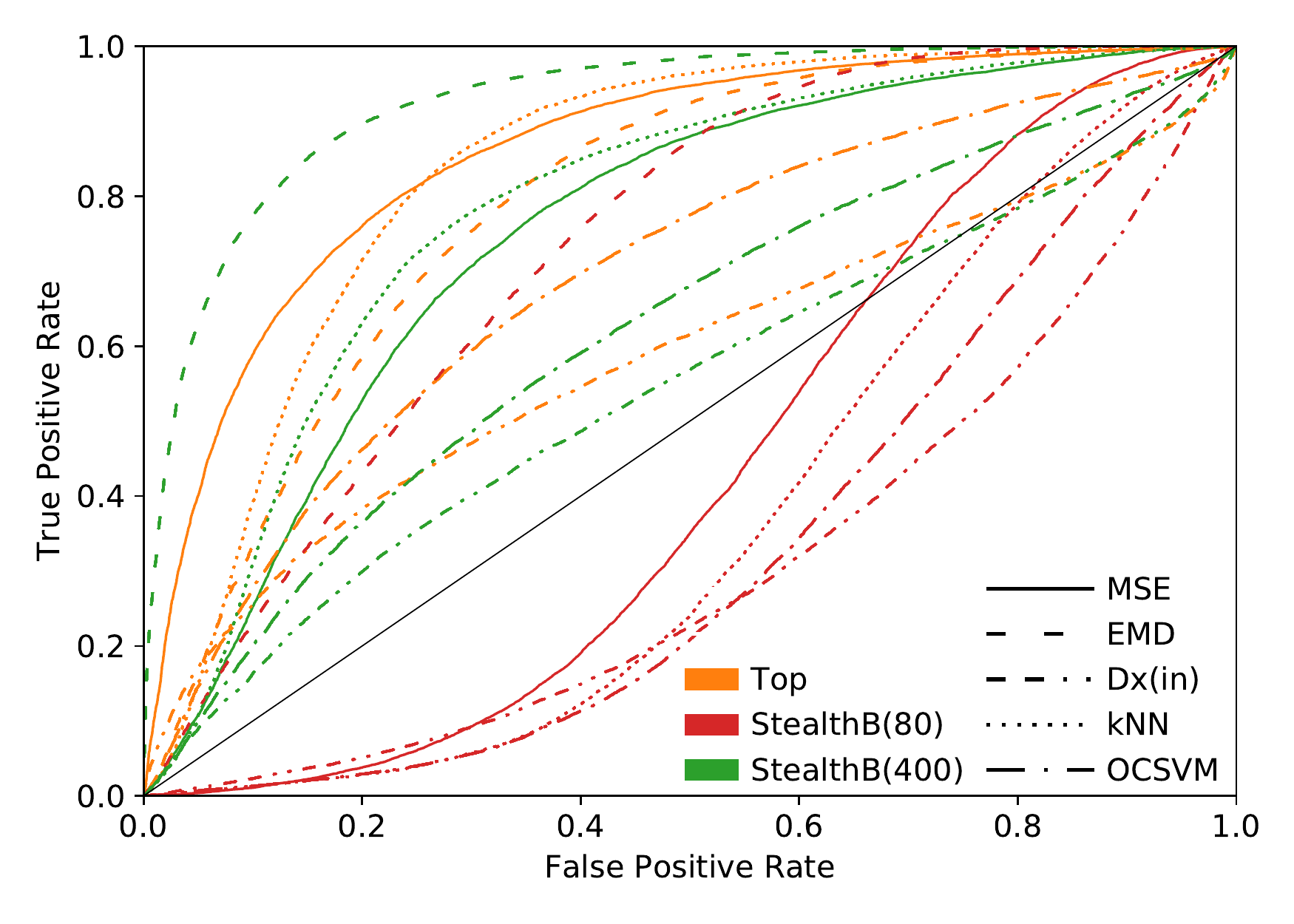}
    \caption{Jet mass of the input and reconstructed jets (left) and ROC curves calculated from the metrics for separating top jets (orange), low mass (red) and high mass (green) Stealth Bosons from QCD jets.}
    \label{fig:performance}
\end{figure}

In the training three loss terms are being balanced and jointly minimised, however one of the key measures of performance is the ability to reconstruct the masses of jets.
In Fig.~\ref{fig:performance} the input and reconstructed jet mass are compared for QCD jets, top jets, and the two Stealth Boson jets. The network is able to reconstruct the mass of QCD jets to a high degree, with the signal samples all being reasonably well modelled. This demonstrates that the network has learned some of the underlying physics, though it is not able to perfectly capture the resonances. 

Furthermore, as the aim is to use the VAE for anomaly detection, the ability to separate the background, QCD, from different signals is therefore also important to measure. The ROC curves for the different anomaly metrics for separating each of the three signal samples from QCD are shown in Fig.~\ref{fig:performance}. Here it can be seen that only the EMD metric can correctly identify StealthB(80) as anomalous, with the other metrics assigning this signal to be more nominal-like than QCD.

In the case of Stealth(80), the jets are from a much narrower resonance than top and StealthB(400) jets. They also have a much higher $p_{\mathrm{T}}$ distribution compared to QCD and top jets. This boosting leads to a very collimated jet, with the opening angle correlated to the ratio of the jet mass and the jet $p_\mathrm{T}$. This effect can be seen in the average jet images in Fig.~\ref{fig:jetimg}, where the StealthB(80) jets have less identifiable substructure, and where the distribution of constituent momentum is strongly concentrated in the centre of the jet. As a result, it is much harder for the current metrics to identify the jets as anomalous.
Excluding StealthB(80), the other metrics offer a range of performance, with the observation space metrics and kNN performing particularly well at identifying top and StealthB(400) jets from QCD.


\section{The Question of Mass}

Just as the jet mass is a key property to study how well known jets are reconstructed, and which is used to validate the autoencoder performance, it is also hugely important in the context of anomaly detection.
When searching for new phenomena in particle physics, the most quantifying property of any observation is the observed mass of a resonance. As such, it is crucial that we are able to identify anomalous jets for a range of mass values, and that when applying a cut on an anomaly metric we do not artificially create a signal from sculpting of the mass distribution.

\begin{figure}[h]
    \centering
    \includegraphics[width=0.45\textwidth]{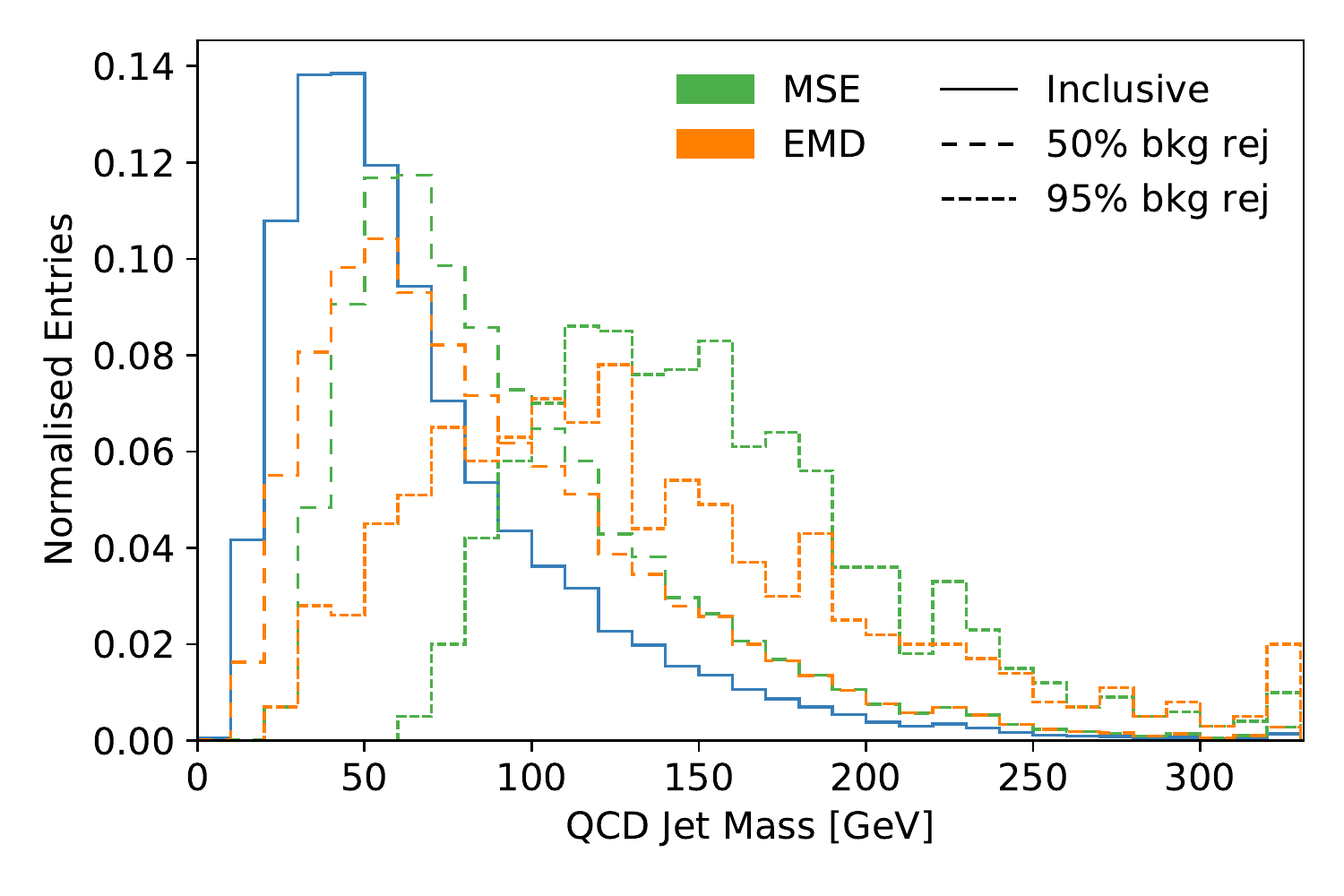}\hskip4ex
    \includegraphics[width=0.45\textwidth]{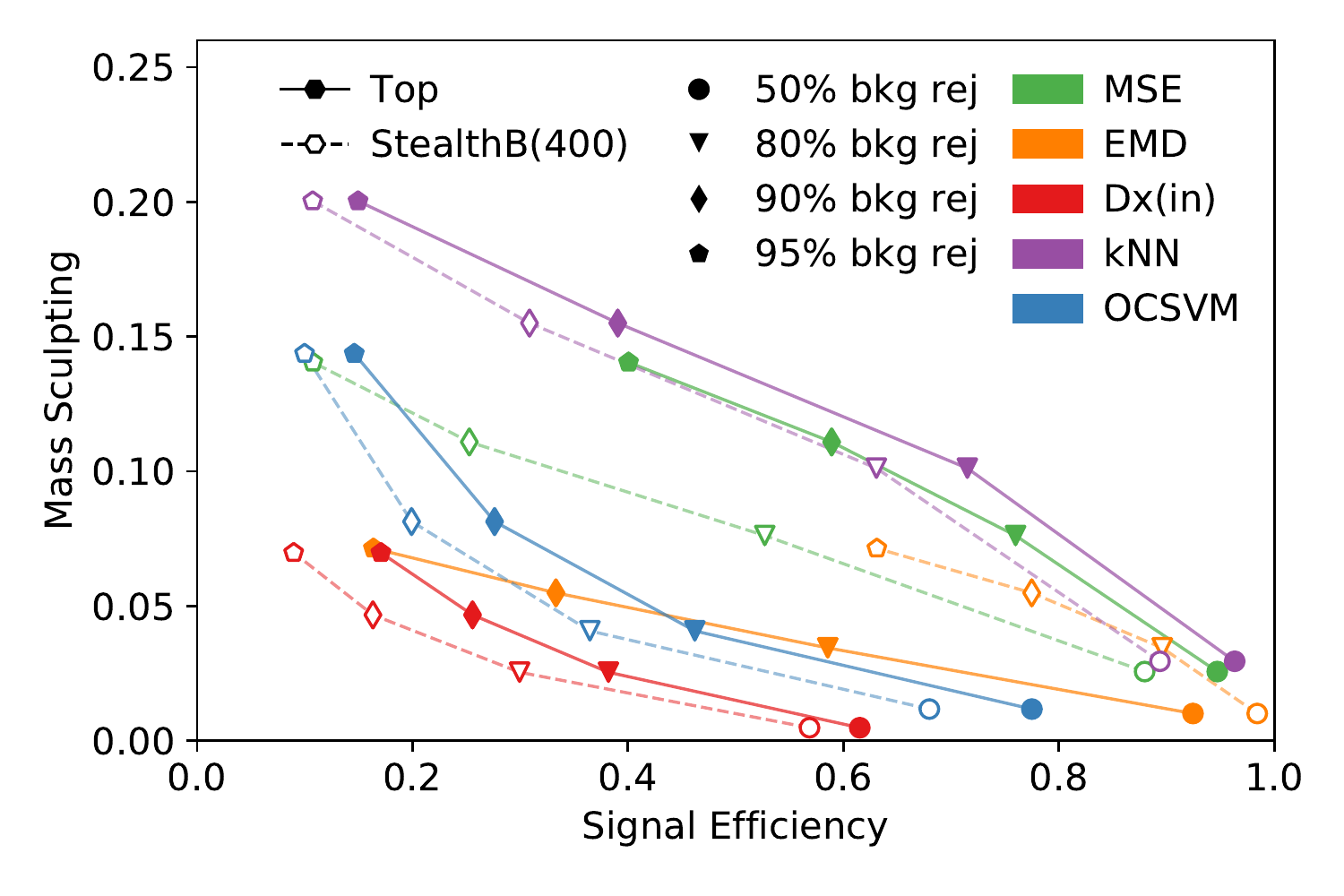}
    \caption{QCD jet mass sculpting after applying cuts on MSE and EMD (left).
    Comparing signal efficiency versus mass sculpting (right) of the metrics for top jets (solid) and StealthB(400) (hollow).
    }
    \label{fig:mass_sculpt}
\end{figure}

As has been seen in the previous section, it is more difficult for the anomaly metrics to identify the lower mass StealthB(80) correctly. Furthermore, when applying a cut on the MSE distribution, we can see that the QCD background is indeed sculpted and biased towards higher mass jets, as shown in Fig.~\ref{fig:mass_sculpt}. To quantify the performance of all the metrics, the amount of mass sculpting is compared to the rejection efficiency of QCD for each of the metrics, alongside the signal acceptance, for top and StealthB(400) jets.
Mass sculpting is measured from the relative entropy of the QCD jet mass distribution after applying the cut to the inclusive distribution.
The ideal metric would have a flat response in sculpting as a function of background rejection, whilst maintaining high signal efficiency, for both signals. Of the metrics studied, kNN is particularly bad for mass sculpting, whereas EMD has the least impact on selected jet masses. 



\begin{figure}[h]
    \centering
    \includegraphics[width=0.45\textwidth]{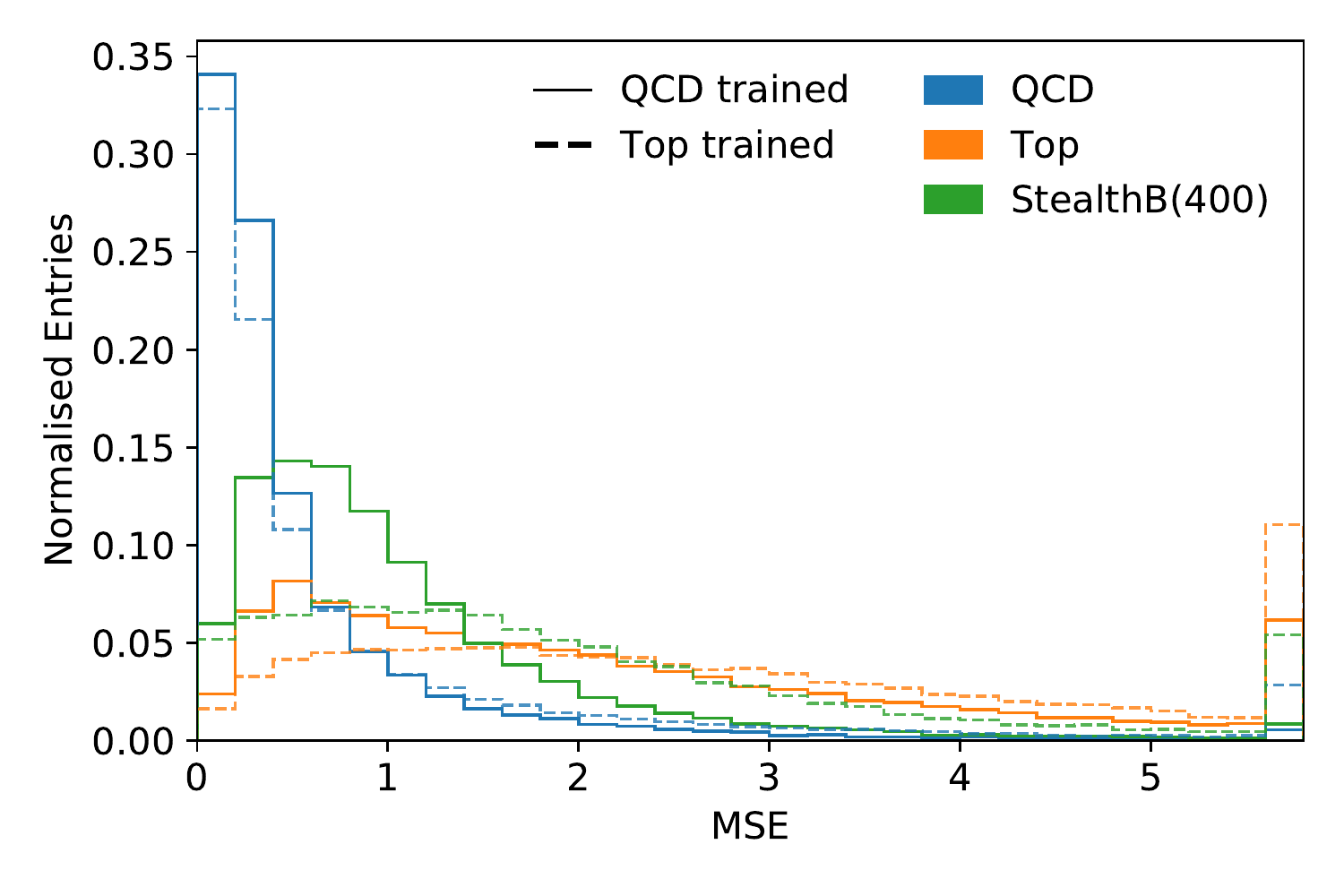}\hskip4ex
    \includegraphics[width=0.45\textwidth]{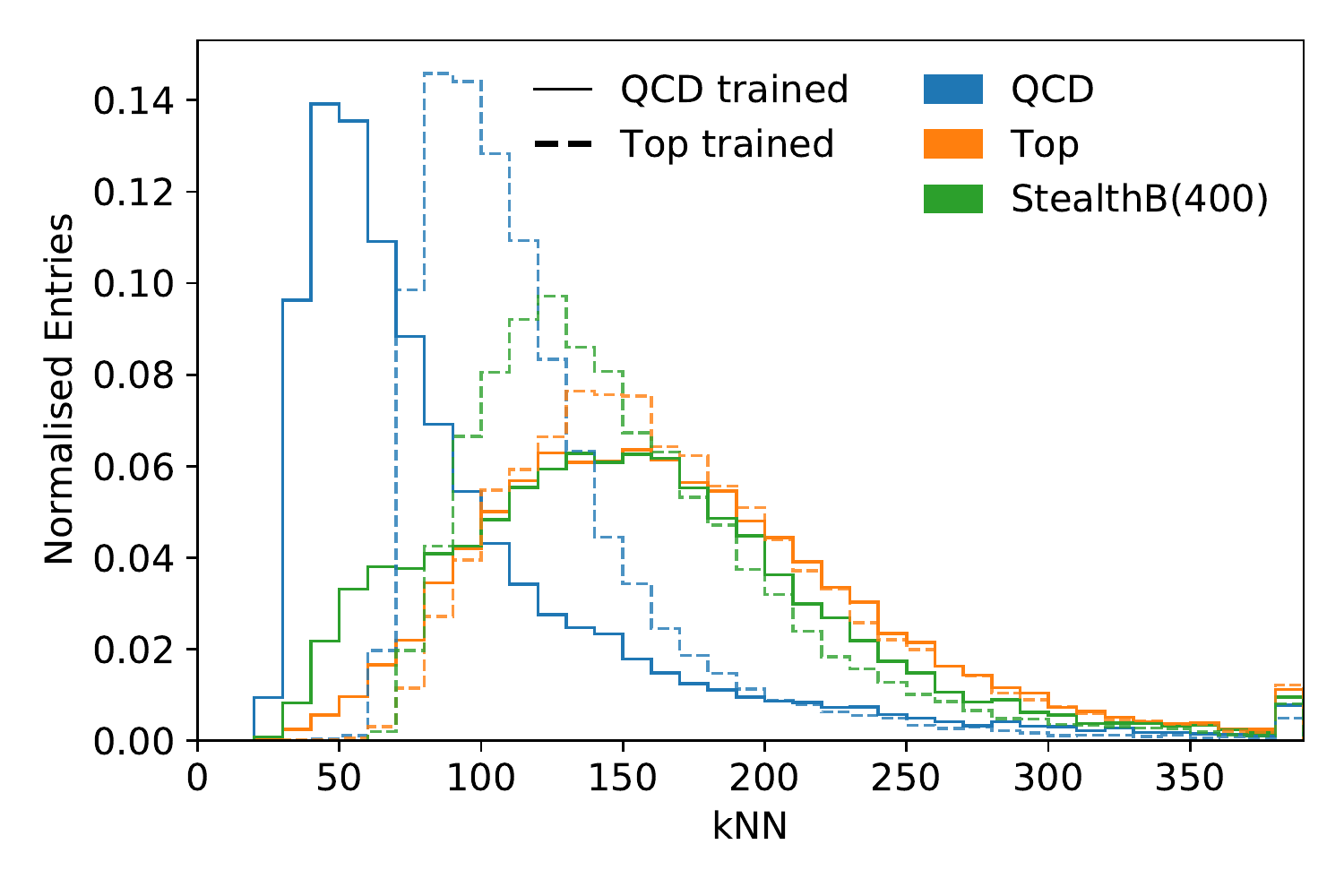}
    \caption{Comparing MSE score (left) and kNN (right) for QCD, top jets, and StealthB(400) when training the model and metrics with QCD (solid) or Top (dashed) as nominal.}
    \label{fig:invert}
\end{figure}

Another robustness test of anomaly metrics is to change the definition of what is nominal, and to train the network using a more massive signal than QCD, for example top jets.
In Fig.~\ref{fig:invert} this comparison has been performed for MSE on the model trained on top and QCD jets, as well as for kNN in the latent space.
In both observation space and latent space metrics, higher mass signals are seen as more anomalous, even when used as the nominal dataset for training. In particular, MSE is almost unchanged for top jets and QCD, regardless of which sample is used for training.
This metric is highly correlated to jet mass, and anti correlated to jet $p_{\mathrm{T}}$. These are two underlying properties which are dominated by the relative position and momenta of the two leading constituents of the jet.

\section{Conclusions}

As can be seen from the studies presented in this paper, particular care needs to be taken when designing anomaly metrics for jets in high energy physics.
Jet mass is a key property in assessing reconstruction performance, but also for the eventual application of anomaly detection in an analysis.
As such, the performance of multiple approaches should be evaluated with a variety of signals at different masses, in order to choose an approach not with the best absolute performance, but which is most robust to signals of all masses and protection against artificially introducing a signal through mass sculpting.

Outlier Exposure presents one avenue of study with promising results, as investigated in Ref.~\cite{oevae}. Furthermore, attempts to decorrelate the anomaly metrics or components of the architecture are another direction of interest. Using techniques to directly remove the correlation to the jet mass from the latent space are under investigation, and making all  constraints $\mathcal{D}_{\mathrm{x}\mathrm{\hat{x}}}$, $\mathcal{D}_{\mathrm{\hat{x}}}$ and $\mathcal{D}_{\mathrm{z}}$ conditionl on the jet mass provides an additional avenue of study.

Further studies should also test the generalisability of techniques to identify anomalous jets from the nominal sample, regardless of what is used in training, ensuring that anomalies are detected using jet substructure, not simply identifying the mass of the jets.
We believe that in all cases, similar benchmark metrics to those presented in this paper will form a necessary component in the performance evaluation.



\clearpage

\section*{Broader Impact}

The proposed approach is based on the information bottleneck formulation applied to anomaly detection in high energy physics. In particular we focus on one of the major complexities involved with applying this family of approaches to jet physics, namely the large and non-trivial dependence on the jet mass.
The approach is generic in nature and subsequently can be used in a wide range of practical applications, and the lessons learned from understanding the relationship between what is seen as anomalous and latent properties is equally transferable to other domains.
In this paper we do not address any ethical aspects or societal consequences, as the focus is on understanding the fundamental nature of the universe.
A positive outcome of this work is in establishing a series of checks and considerations to enable a fair comparison of approaches and benchmarks. No negative outcomes are foreseen as this paper seeks to promote engagement in the community and help improve theoretical and empirical understanding of this burgeoning field of research. Though we cannot at this time present the holy grail that is a solution to this problem in a fully unsupervised setting, this is the subject of our future work. We also plan to incorporate into a full statistical analysis, searching for signs of new elementary particles in proton collisions, and devise a generic approach that would have the potential to be applied to other domains.





\end{document}